\def\bold#1{\setbox0=\hbox{$#1$}%
     \kern-.025em\copy0\kern-\wd0
     \kern.05em\copy0\kern-\wd0
     \kern-.025em\raise.0433em\box0 }
\def\slash#1{\setbox0=\hbox{$#1$}#1\hskip-\wd0\dimen0=5pt\advance
       \dimen0 by-\ht0\advance\dimen0 by\dp0\lower0.5\dimen0\hbox
         to\wd0{\hss\sl/\/\hss}}
\documentstyle[12pt]{article}
\renewcommand{\topfraction}{1.0}
\renewcommand{\bottomfraction}{1.0}
\renewcommand{\textfraction}{0.0}
\newlength{\dinwidth}
\newlength{\dinmargin}
\newcommand{\resection}[1]{\setcounter{equation}{0}\section{#1}}

\begin{document}

\def\lq{\left [}
\def\rq{\right ]}
\def\LL{{\cal L}}
\def\VV{{\cal V}}
\def\AA{{\cal A}}

\newcommand{\be}{\begin{equation}}
\newcommand{\ee}{\end{equation}}
\newcommand{\bea}{\begin{eqnarray}}
\newcommand{\eea}{\end{eqnarray}}
\newcommand{\nn}{\nonumber}
\newcommand{\dd}{\displaystyle}
\def\e{{\rm e}}
\thispagestyle{empty}
\vspace*{4cm}
\begin{center}
  \begin{Large}
  \begin{bf}
Penguin contributions to rates and CP asymmetries in non-leptonic B-decays.
Possible experimental procedures and estimates.\\
  \end{bf}
  \end{Large}
  \vspace{5mm}
  \begin{large}
A. Deandrea, N. Di Bartolomeo and R. Gatto\\
  \end{large}
D\'epartement de Physique Th\'eorique, Univ. de Gen\`eve\\
  \vspace{5mm}
  \begin{large}
F. Feruglio\\
  \end{large}
Dipartimento di Fisica, Univ. di Padova\\
  \vspace{5mm}
  \begin{large}
G. Nardulli\\
  \end{large}
Dipartimento di Fisica, Univ.di Bari\\
I.N.F.N., Sezione di Bari\\
  \vspace{5mm}
\end{center}
  \vspace{2cm}
\begin{center}
UGVA-DPT 1993/10-837\\\
hep-ph/9310326\\\
October 1993
\end{center}
\vspace{1cm}
\noindent
$^*$ Partially supported by the Swiss National  Science Foundation
\newpage
\thispagestyle{empty}
\begin{quotation}
\vspace*{5cm}
\begin{center}
  \begin{Large}
  \begin{bf}
  ABSTRACT
  \end{bf}
  \end{Large}
\end{center}
  \vspace{5mm}
\noindent
We present a study of non-leptonic B-meson decays, partial widths and CP
asymmetries, and discuss the possibilities of determining the penguin
contributions through less model-dependent phenomenological analyses. In the
last section we employ a specific model to give more definite numerical
predictions.
\end{quotation}
\newpage
\setcounter{page}{1}
\setcounter{section}{0}
\section{Introduction and summary of content}

In non-leptonic decays of B-mesons the penguin contributions of the effective
$\Delta B=1$ hamiltonian become non-negligible when the tree contributions are
suppressed and they are even dominant in certain cases.
Moreover for those $B$ decays which could give rise to observable CP
asymmetries, the penguin contributions might cause large deviations from the
tree predictions \cite{Gronau}, even when numerically much smaller than the
latter. In such cases a correct estimate of the penguin effects is therefore
crucial.
 We present in this note
a study of such decays.
We will discuss the role of the penguin amplitudes both in the case where they
are not negligible in the computation of the widths and in the case where,
although not important to evaluate the decay widths, they are relevant to the
analysis of CP asymmetries.
     At least in principle, the experimental informations on
the first set of decays (essentially those corresponding to the
Cabibbo suppressed $b->u{\bar u} s$ transitions) might provide
bounds, constraints or even a direct access to those parameters
defining the penguin contributions to the CP violating asymmetries.
We shall stress as much as possible procedures which are less model-dependent.

The calculations make use of the recently developed two-loop effective $\Delta
B=1$
hamiltonian \cite{Buras} and of the factorization approximation. The form
of the amplitudes are collected in Tables I, II.
They depend on the Cabibbo-Kobayashi-Maskawa matrix elements
$V_{i,j}$, on combinations $a_i$ of the Wilson coefficients
of the effective hamiltonian and on the hadronic matrix elements
of the weak currents.
 In view of the inevitable
theoretical uncertainties we suggest the experimentalists to mostly rely
directly on the parametrized expressions in Tables I and II to fit the
theoretical parameters from the ratios of the partial widths which are or
 will eventually become available (we give some example).

The asymmetries for decays of neutral B-mesons into states with
CP self-conjugate particle composition are given in Tables III. The usual
asymmetry parameter (neglecting strong phases) $Im (q\bar{A}/pA)$ depends on
the mixing phase $\varphi_M$ and on the amplitudes $A_T \e^{i\varphi_T}$ and
$A_P \e^{i\varphi_P}$ for tree and penguin respectively. The angles
$\varphi_T$, $\varphi_P$, and the expressions for the ratios $A_P/A_T$ are
reported in Tables III. We recall that in some cases the rate asymmetry is
reduced with respect to the fundamental asymmetry.

As mentioned above,
Tables III, in conjunction with the previous tables I and II, can be used
to extract, in a way which is as much as possible model
independent, the relevant parameters, by forming suitable ratios of partial
rates and estimating or bounding the asymmetries from experimental data
directly. We illustrate such a procedure with three examples, but various
possibilities may appear and the procedures finally chosen will presumably
depend on the kind of experimental data that will become available at the time
of the analysis.

The full predictivity of the model that had been developed in ref.
\cite{noi1} will be exploited in the last part of this note to get at the
numbers presented in Tables IV.
We are mainly interested in obtaining an order-of-magnitude
estimate of the rates for the processes considered in section 2,
just to have a rough idea of the rareness of the considered
decay modes and of their observability at future dedicated facilities.
Indeed, as explained above and shown in detail in section 2,
we think that, maybe, the best use of the these channels
could consist, via the study of appropriate ratios of rates,
in constraining the parameters entering the theoretical
analysis via the factorization hypothesis.
 These predictions should be taken with care,
as they are subject to the details of the particular model and to the present
inputs. The theoretical errors are difficult to estimate. In some case their
instability with respect to input parameters (for instance $N_c$) suggests that
they should be taken with great caution.

A very large literature already exists on B-meson decays. We shall refer to
a recent book edited by Stone \cite{prSto}, and for CP violation to the
books edited by Wolfenstein \cite{prWo} and by C.~Jarlskog \cite{prJa}.

\resection{Parametrization of the amplitudes}

The two-loop effective hamiltonian  for $\Delta B=1$ transitions can be written
as follows \cite{Buras}:
\bea
H_{NL}=\frac{G_F}{\sqrt{2}} \left[ V_{ub}V^*_{uq} (c_1 O_1^u+c_2 O_2^u)
\right. & + &
V_{cb}V^*_{cq} (c_1 O_1^c + c_2 O_2^c)+  \nn \\
 & -& \left. V_{tb}V^*_{tq} \sum^6_{i=3} c_i O_i \right] +h.c.
\label{1}
\eea
where $q=d,s$, $c_i$ are the Wilson coefficients at the scale $\mu \approx
m_b$; $O_1^{u,c}$, $O_2^{u,c}$ are the operators:
\bea
O_1^u=({\bar u}b)_{V-A} ({\bar q}u)_{V-A} \, \, &\,& \,\,
O_1^c=({\bar c}b)_{V-A} ({\bar q}c)_{V-A}  \nn \\
O_2^u=({\bar q}b)_{V-A} ({\bar u}u)_{V-A} \, \, &\,& \,\,
O_2^c=({\bar q}b)_{V-A} ({\bar c}c)_{V-A}
\label{2}
\eea
$O_i$ ($i=3, \ldots 6$) are the so-called penguin operators
\bea
O_3 &=&({\bar q}b)_{V-A} \sum_{q'} ({\bar q}'q')_{V-A} \nn \\
O_4 &=& \sum_{q'}({\bar q}'b)_{V-A} ({\bar q}q')_{V-A} \nn \\
O_5 &=&({\bar q}b)_{V-A} \sum_{q'} ({\bar q}'q')_{V+A} \nn \\
O_6 &=& -2 \sum_{q'}({\bar q}'(1-\gamma_5)b)
({\bar q}(1+\gamma_5)q')
\label{3}
\eea
In the previous formulae $q=d,s$ and $({\bar u}b)_{V-A}={\bar
u}\gamma^\mu(1-\gamma_5)b$.
The values of the Wilson coefficients, for $m_b=4.8 GeV$, $m_{top}=150~GeV$ and
$\Lambda_{{\bar MS}}=250~MeV$ are \cite{Buras}
\bea
c_1 &=& 1.133, \; c_2=-0.291, \; c_3=0.015 \nn \\
c_4 &=& -0.034, \; c_5=0.010,  \; c_6=-0.042
\label{4}
\eea

The hamiltonian \ref{1} allows to study non leptonic decays of $B$ mesons. In a
previous paper \cite{noi}, using the factorization approach, we focused on
processes where only the operators $O_1$ and $O_2$ were relevant, at least for
what concerns the rates. Notice that the penguin operators have Wilson
coefficients very small as compared to those of $O_1$ and $O_2$, but they
become
important when the tree diagram is strongly Cabibbo suppressed or when the
decay can only occur through penguin diagrams. Here we will be interested in
such processes. Moreover we will study the influence of the penguins on CP
asymmetries for neutral $B$ mesons.

Let us examine first the Cabibbo suppressed decays; in calculating the rates it
turns out that the penguin contribution is relevant  in the flavour processes
$b \to u {\bar u}s$: in this case the tree CKM factor is $V_{ub}V^*_{us}$,
of the order $\lambda^4$ ($\lambda$ is the Cabibbo angle),
 while the penguin has the factor $V_{tb} V^*_{ts}$,
of the order $\lambda^2$, which compensates the smallness of the Wilson
coefficients. In Table I we give the amplitudes for all these processes,
calculated using factorization and the following parametrization of the
hadronic matrix elements
\be
<P(p')|V^{\mu}|B(p)> =
 \big[ (p+p')^{\mu}+\frac{M_P^2-M_B^2}{q^2} q^{\mu}\big]
F_1(q^2) -\frac {M_P^2-M_B^2}{q^2} q^{\mu} F_0(q^2)
\label{5}
\ee
\bea
<V (\epsilon,p')| &(&V^{\mu}-A^{\mu})|B(p)> =
\frac {2 V(q^2)} {M_B+M_V}
\epsilon^{\mu \nu \alpha \beta}\epsilon^*_{\nu} p_{\alpha} p'_{\beta} \nn\\
&+& i  (M_B+M_V)\left[ \epsilon^*_\mu -\frac{\epsilon^* \cdot q}{q^2}
q_\mu \right] A_1 (q^2) \nn\\
& - & i \frac{\epsilon^* \cdot q}{(M_B+M_V)} \left[ (p+p')_\mu -
\frac{M_B^2-M_V^2}{q^2} q_\mu \right] A_2 (q^2) \nn \\
& + & i \epsilon^* \cdot q \frac{2 M_V}{q^2} q_\mu A_0 (q^2)
\label{6}
\eea
where $P$ is a  pseudoscalar meson and $V$ a  vector meson, $q=p-p'$,
\be
A_0 (0)=\frac {M_V-M_B} {2M_V} A_2(0) + \frac {M_V+M_B}
{2M_V} A_1(0)
\label{7}
\ee
and $F_1 (0)=F_0 (0)$.
We take also:
\bea
<0|A^\mu|P(p)> &=& -i f_P p^\mu \nn \\
<0|V^\mu|V(p, \epsilon )> &=&  f_V M_V \epsilon^\mu
\label{8}
\eea
The coefficients $a_i$ ($i=1, \ldots 6$) are given by
\be
a_{2i-1}=c_{2i-1}+\frac{c_{2i}}{N_c} \; \; \;
a_{2i}=c_{2i}+\frac{c_{2i-1}}{N_c} \; \; \; i=1,2,3
\label{9}
\ee
where in the naive factorization
 $N_c$ is the number of colors, but due to the uncertainties of the approach
it is treated as a parameter to be determined by the data. For example in $D$
decays  the value $N_c=\infty$ agrees with the experiments.
 In Table II we give  the amplitudes in the factorization approximation for
the pure penguin decays of ${\bar B}^0$, ${\bar B}_s$ and $B^-$ mesons.

A possible experimental determination of the coefficients $a_i$ could be done
through ratios of widths, such that most of the hadronic uncertainties cancel
out. As an example one could measure $a_4$ from the ratio:
\be
R=\frac{\Gamma({\bar B}^0 \to \pi^+ K^{*-})}{\Gamma ({\bar B}^0 \to \pi^+
\rho^-)} =\left( \frac{f_{K^*}}{f_{\rho}} \right)^2 \frac{|V_{ub}V^*_{us} a_1-
V_{tb}V^*_{ts} a_4|^2}{|V_{ub}V^*_{ud} a_1|^2}
\label{10}
\ee
We remark that the width for ${\bar B}^0 \to \pi^+ K^{*-}$ depends on the
relative phase between tree and penguin contributions; in general one should
expect a strong and a weak phase, that in this case is $\gamma$. As for the
strong phases, they are expected to decrease as the inverse heavy meson mass
and therefore we neglect them.

\resection{CP asymmetries}

The penguin amplitude, through interference with the tree amplitude,
contributes to CP asymmetries in $B$ decays . Neglecting possible
strong phases, we write the amplitude of the process as a sum of a penguin
and a tree contribution \cite{Gronau}:
\be
A=A_T \e^{i\varphi_T} +A_P \e^{i\varphi_P}
\label{11}
\ee
where $\varphi_{T,P}$ are the weak phases. The time dependent asymmetry of a
neutral $B$ meson ($B^0$ or $B_s$) decaying into a CP-eigenstate can be
written as
\be
a(t)=-Im \left(\frac{q}{p} \frac{{\bar A}}{A} \right) \sin (\Delta m t)
\label{12}
\ee
where
\be
A=<f|B^0> \; \; \; {\bar A}=<f|{\bar B}^0>
\label{12bis}
\ee
 $q/p$ is the mixing and $\Delta m$ the mass difference of the two
neutral $B$ mesons. If $A_P/A_T<<1$ we can make an expansion, obtaining:
\be
Im  \left(\frac{q}{p} \frac{{\bar A}}{A} \right)= -\sin 2(\varphi_M+\varphi_T)
-
2 \frac{A_P}{A_T} \cos 2(\varphi_M+\varphi_T) \sin (\varphi_P-\varphi_T)
\label{13}
\ee
where $\varphi_M$ is the mixing phase, defined as
\be
\frac{q}{p}=\e^{-2i\varphi_M}
\label{14}
\ee
If $\varphi_T=\varphi_P$, the penguin does not influence the asymmetry, as it
happens
in $B^0 \to \psi K_s$: but in general one has a phase difference. Notice that
the relative contribution of the penguin in the asymmetry depends on the values
of the angles $\varphi_M$, $\varphi_T$ and $\varphi_P$:
 even when $A_P/A_T <<1$ it could
be quite sizeable (or dominating). In Table III we give, for many ${\bar B}^0$
and ${\bar B}_s$ decays into states with CP self-conjugate particle
composition,
the values of the phases $\varphi_T$ and $\varphi_P$;
 we give also the branching
ratios as in ref.\cite{noi}, and the ratio $A_P/A_T$ as a function of the
coefficients $a_i$, assuming factorization.
 The angle $\alpha$, $\beta$, $\gamma$ are
the usual ones, and $\beta '$ is defined as:
\be
\beta '= arg \left[ -\frac{V_{cs}V^*_{cb}}{V_{ts}V^*_{tb}} \right]
\label{15}
\ee
This angle is very small, $|\sin 2\beta'|<0.06$.
The knowledge of the values of the coefficients $a_i$ allows to evaluate the
ratio $A_P/A_T$; in ref. \cite{Wolf} this one is extracted, for the decay
$B^0 \to \pi^+ \pi^-$, from the ratio of the widths  of $\pi^+ K^-$   to
$\pi^+ \pi^-$. As discussed in the previous section the advantage of making
ratios is the cancellation of some hadronic uncertainties.
As examples we will consider the decays ${\bar B}_s \to D_s^+ D_s^-$,
${\bar B}_s \to \rho^0 K_s$ and ${\bar B}_s \to K^+ K^-$.
 Other examples can be constructed using our tables.
The final choices will depend on the relative quality of the experimental
information that will become available.

The amplitude for the process ${\bar B}_s \to D_s^+ D_s^-$ is
\be
{\bar A}= A_T \e^{i\beta'}+A_P
\label{15bis}
\ee
where
\bea
A_T &=& \frac{G_F}{\sqrt{2}} |V_{cb} V^*_{cs}| a_1 <D^+_s|\bar{c}b_-|\bar{B}_s>
<D^-_s|\bar{s}c_-|0> \nn \\
A_P &=& \frac{G_F}{\sqrt{2}} |V_{tb} V^*_{ts}| \left(a_4+2a_6\frac{M^2_{D_s}}
{(m_b-m_c) (m_c+m_s)}\right) <D^+_s|\bar{c}b_-|\bar{B}_s>
<D^-_s|\bar{s}c_-|0>
\label{16}
\eea
and ${\bar q}q'_-={\bar q}\gamma^\mu (1-\gamma_5)q'$.
{}From \ref{15bis} we obtain:
\be
Im \left(\frac{q}{p}\frac{\bar{A}}{A}\right)= \frac{\sin (2\beta') +2 x
\sin\beta'}{1+x^2+2x\cos\beta'}\simeq \frac{2\beta'}{1+x}
\label{17}
\ee
where we have used the smallness of $\beta '$ ($\beta ' \leq 0.03$). The
asymmetry is given by \ref{12}. The value
of $x=A_P/A_T$ can be extracted from the ratio of the widths
\bea
R &= &\frac{\Gamma (\bar{B}_s \to D^+_s D^-_s)}
{\Gamma (\bar{B}_s \to D^+_s \pi^-)}
= \dd \left(\frac{f_{D_s}}{f_{\pi}} \right)^2\frac{|\vec{p}_{D_s}|}
{|\vec{p}_{\pi}|} \left(\frac{F_0^{B_s \to D_s} (M^2_{D_s})}
{F_0^{B_s \to D_s} (M^2_{\pi})} \right)^2 (1+x^2)\nn \\
&=&\dd \left(\frac{f_{D_s}}{f_{\pi}} \right)^2\frac{|\vec{p}_{D_s}|}
{|\vec{p}_{\pi}|} \left[\frac{(M_{B_s}+M_{D_s})^2-M^2_{D_s}}
{(M_{B_s}+M_{D_s})^2-M^2_{\pi}}\right]^2 \left[\frac{\xi (v\cdot v')}
{\xi (v\cdot v'')} \right]^2 (1+x^2 )
\label{18}
\eea
where $\xi$ is the Isgur-Wise form factor, $v\cdot v' =M_{B_s}/(2 M_{D_s})$
and
$v\cdot v'' =(M^2_{B_s}+M^2_{D_s}-M^2_{\pi})/(2M_{B_s} M_{D_s})$. We work in
the infinity quark mass limit.

An interesting decay for the measure of $\gamma$ is $\bar{B}_s \to \rho^0 K_s$,
whose amplitude can be written as
\be
\bar{A} = A_T \e^{-i\gamma}+A_P \e^{i\beta}
\ee
\bea
A_T&=\dd \frac{G_F}{\sqrt{2}} |V_{ub} V^*_{ud}| a_2 <K_s|\bar{d}b_-|\bar{B}_s>
<\rho^0|\bar{u}u_-|0>\nn \\
A_P&= \dd \frac{G_F}{\sqrt{2}} |V_{tb} V^*_{td}| a_4 <K_s|\bar{d}b_-|\bar{B}_s>
<\rho^0|\bar{u}u_-|0>
\label{19}
\eea
For the CP violating asymmetry for the decay $\bar{B}_s \to \rho^0 K_s$
one has
\be
Im \left(\frac{q}{p}\frac{\bar{A}}{A}\right)= \frac{-\sin (2\gamma) +2 x
\sin(\beta-\gamma)+x^2\sin(2\beta)}{1+x^2+2x\cos(\beta+\gamma)}
\label{20}
\ee
For $\bar{B}_s \to \rho^- K^+$ we have
\be
A = \sqrt{2} \left(\frac{a_1 A_T}{a_2} \e^{-i\gamma}-A_P \e^{i\beta}
\right)
\ee
The ratio of the two widths gives
\be
R=\frac{\Gamma (\bar{B}_s \to \rho^0 K^0)}{\Gamma (\bar{B}_s \to \rho^- K^+)}
= \frac{1}{2} \frac{\left[ 1+x^2+2x\cos (\beta +\gamma ) \right]}{\left[
1+a_2^2 x^2/a_1^2 -2x a_2/a_1 \cos (\beta +\gamma)\right]}
\label{21}
\ee
Therefore for each given value of $\beta$ and $R$ we can extract
 $x=A_P/A_T$ as a
function of $\gamma$; from \ref{20} we can after that obtain $\gamma$, once
the asymmetry, given by \ref{12}, is measured. The ratio $a_2/a_1$, with its
sign, is obtained by experimental results as for example $BR(B^- \to D^0
\pi^-)/BR({\bar B}^0 \to D^+ \pi^-)$; for instance in \cite{noi} we find
$a_2/a_1 \simeq +0.27$.

Another decay potentially interesting to determine $\gamma$ is ${\bar B}_s \to
K^+K^-$. The amplitude is:
\be
\bar{A} = A_T \e^{-i\gamma}+A_P
\ee
\bea
A_T&=&\dd \frac{G_F}{\sqrt{2}} |V_{ub} V^*_{us}| a_1 <K^+|\bar{u}b_-|\bar{B}_s>
<K^-|\bar{s}u_-|0>\nn \\
A_P&=& \dd \frac{G_F}{\sqrt{2}} |V_{tb} V^*_{ts}| \left( a_4 +\frac{2 a_6
M_K^2}{(m_b-m_u)(m_u+m_s)} \right)
 <K^+|\bar{u}b_-|\bar{B}_s> <K^-|\bar{s}u_-|0>
\label{22}
\eea
Notice that in this case $A_P$ and $A_T$ are of the same order.
The asymmetry is
proportional to:
\be
Im \left(\frac{q}{p}\frac{\bar{A}}{A}\right)= -\frac{\sin (2\gamma) +2 x
\sin\gamma}{1+x^2+2x\cos\gamma}
\label{23}
\ee
Similarly to the previous case one can extract the value of $x$ from the
following ratio
\be
R=\frac{\Gamma (\bar{B}_s \to K^+ K^-)}{\Gamma (\bar{B}_s \to K^+ \pi^-)}
=\left( \frac{f_K}{f_{\pi}}\right)^2
\frac{\lambda^2\left[ 1+x^2+2x\cos\gamma  \right]}{\left[
1+x^2 y^2 \lambda^4 -2x  y \lambda^2 \cos (\beta +\gamma)\right]}
\label{24}
\ee
where
\be
y=\frac{\tan\beta}{\tan\beta\cos\gamma+\sin\gamma}
\label{25}
\ee
As for the ratio \ref{21} knowing $\beta$, $R$ and the asymmetry \ref{23}
one can extract the value of $\gamma$.

\resection{Numerical results}

So far we have not exploited the model developed in ref. \cite{noi1} to
attempt at more complete (and also more risky) numerical estimates. We
will do this in the present section and in Tables IV. The numerical
evaluation of the rates requires the knowledge of the form factors.
Moreover one has to chose a value for the parameter $N_c$ initially introduced
as giving the number of colors, but usually assumed as a phenomenological
parameter to account for the uncertainties of the model. The analysis
done in \cite{noi} seems to indicate $N_c \approx 2$ as an effective value,
but the problem is open and the theoretical approach still to be clarified.
We notice that especially $a_2$, $a_3$ and $a_5$ are strongly
dependent on $N_c$ and the rates dominated by these coefficients can have
large variations.

The results obtained with the model \cite{noi1} and $N_c=2$ are reported in
Table IVa and IVb, respectively for the decay of Table I and II. In Table IVa
we have simply marked with a star (*) the strongly $N_c$ dependent decays,
while in Table IVb we have also quoted the values with $N_c=3$ and
$N_c=\infty$, because of the large variations of the branching ratio with
$N_c$. We notice that for all the decays of Table
I the widths have an interference term between penguin and tree depending on
$\cos \gamma$, which at present is unconstrained. Therefore in Table IVa we
present the branching ratios for the two extreme values $\cos \gamma=\pm 1$.

Experimentally only upper limits exist at present for the processes in Table
IV, and our values are below these data. The most stringent bounds as compared
with our results are \cite{PDG} for $Br(B^- \to \phi K^- < 8\times 10^{-5})$
and $Br(\bar{B}^0 \to \pi^+ K^- < 9\times 10^{-5})$. The measure of the rate
$B^- \to \phi K^-$, a pure penguin process, will allow to extract the value of
$|a_3+a_4+a_5|$, which is quite sensitive to $N_c$ (see Table IVb); the rate
for $\bar{B}^0 \to \pi^+ K^-$ depends on the value of $\cos \gamma$ (see Table
IVa), therefore a measure of this width will allow a determination of $\gamma$,
once known the ratio of penguin to tree contribution, which can for example
be measured in the way suggested in \cite{Wolf}, from the ratio
$\bar{B}^0 \to \pi^+\pi^-$ to $\bar{B}^0 \to \pi^+ K^-$. The measure of the
rate for ${\bar B}^0 \to \pi^+ K^-$ could be an
alternative way to determine $\gamma$ with respect to the asymmetry in $B_s$
decays, which presents an experimental challenge and needs time-dependent
measurements.

\newpage
\begin{center}
  \begin{Large}
  \begin{bf}
  Tables Captions
  \end{bf}
  \end{Large}
\end{center}
  \vspace{5mm}
\begin{description}

\item[ Table I] Amplitudes for non-leptonic decays with tree level contribution
strongly Cabibbo suppressed ($V_{ub} V^*_{us}$).

\item[ Table II] Amplitudes for non-leptonic decays that take place only
through
penguin contributions.

\item[ Table III] $\bar{B}_s$ and $\bar{B}^0$ decays into states with
CP self-conjugate particle pairs. The tree level contribution
dominates the width. Tree and penguin weak phases $\varphi_T$ and $\varphi_P$
are given together with the rate of the respective amplitudes.

\item[ Table IVa] Branching ratios for non-leptonic decays with
tree level contribution strongly Cabibbo suppressed ($V_{ub} V^*_{us}$) are
given for two values of the weak angle $\gamma$. The contribution is the same
in the two cases when there is no penguin contribution to the amplitude.
We take $|V_{ub}|=0.03$, $|V_{cb}|=|V_{ts}|=0.042$ and
$\tau_B=\tau_{B_s}=14\times 10^{-13} s$, $m_b=5000 ~MeV$, $m_s=150~MeV$,
$m_d=10~MeV$ and $m_u=5~MeV$. Furthermore we take $f_{\eta}=f_{\pi}$.
We fixed the value $N_c=2$, signaling with (*) the cases where there is a
strong $N_c$ dependence of the result.

\item[ TableIVb] Branching ratios within the model of \cite{noi1} for
non-leptonic decays that take place
only through penguin contributions. As an example of the variation of the
result with $N_c$ we have listed the three cases $N_C=2, 3, \infty$. We take
$V_{td}=0.01$.

\end{description}

\end{document}